\documentclass[showpacs,showkeys]{revtex4}

\usepackage{graphicx}
\usepackage{amsmath}
\usepackage{amssymb}
\usepackage{color}
\begin{document}

\title{Nonassociativity, supersymmetry, and hidden variables}
\author{Vladimir Dzhunushaliev
\footnote{Senior Associate of the Abdus Salam ICTP}}
\email{dzhun@krsu.edu.kg} \affiliation{Dept. Phys. and Microel.
Engineer., Kyrgyz-Russian Slavic University, Bishkek, Kievskaya Str.
44, 720021, Kyrgyz Republic}

\date{\today}

\begin{abstract}
It is shown that the supersymmetric quantum mechanics has an octonionic generalization. The generalization is based on the inclusion of quaternions into octonions. The elements from the coset octonions/quaternions are unobservables bacause they cannot be considered as quantum operators as a consequence of their non-associative properties. The idea that the octonionic generalization of the supersymmetric quantum mechanics describes an observable particle formed with unobservable ``particles'' is presented.

\end{abstract}


\maketitle

\section{Introduction}

Non-associative algebras may be surely called beautiful mathematical entities. Nevertheless, they have never been systematically utilized in physics in any fundamental fashion, although some attempts have been made toward this goal. However, it is still possible that nonassociative algebras may play some essential future role in the ultimate  theory, which is yet to be discovered.
\par
Octonions are one example of a nonassociative algebra. The octonions are the largest normed algebra after the algebras of real numbers, complex numbers, and quaternions \cite{Sc}. Since their discovery in 1844-1845 by Graves and Cayley there have been various attempts to find appropriate uses for octonions in physics (see reviews in Ref's \cite{Oct}, \cite{emch}).
\par
In this paper we would like to show that the supersymmetric quantum mechanics has an octonionic generalization. The generalization is based on the inclusion of quaternions $\mathbb H$ into octonions $\mathbb O, \mathbb H \subset \mathbb O$. The elements from the coset
$\mathbb O / \mathbb H$ can be considered as unobservables bacause they cannot be considered as quantum operators as a consequence of their nonassociative properties.
\par
The paper is organized as follows in sections \ref{split} and \ref{supersymmetry} the introduction to the split octonion algebra and supersymmetric quantum mechanics is given. In section \ref{extension} we present an octonionic generalization of supersymmetric quantum mechanics. In section \ref{hidden} we show that the elements from the coset
$\mathbb O / \mathbb H$ can be considered as unobservables. In section \ref{application} we consider the possible applications of the octonionic generalization of supersymmetric quantum mechanics. In section \ref{conclusion} we present comments and conclusions.

\section{The split octonion algebra}
\label{split}

In this section we follow Ref.~\cite{Gunaydin:1973rs}. A composition algebra is defined as an algebra $A$ with identity and with a nondegenerate quadratic form $Q$ defined over it such that $Q$ permits the composition
\begin{equation}
	Q(xy) = Q(x) Q(y), \; x,y \in A.
\label{1-10}
\end{equation}
According to the Hurwitz theorem, only four different composition algebras exist over the real or complex number fields. These are the real numbers $\mathbb{R}$ of dimension 1, complex numbers $\mathbb{C}$ of dimension 2, quaternions $\mathbb{H}$ of dimension 4, and octonions $\mathbb{O}$ of dimension 8. Of these algebras, the quaternions $\mathbb{H}$ are not commutative and the octonions $\mathbb{O}$ are neither commutative nor associative. A composition algebra is said to be a division algebra if the quadratic form $Q$ has the following property
\begin{equation}
	\text{if } Q(x) = 0 \text{ implies that } x=0.
\label{1-20}
\end{equation}
Otherwise, the algebra is called \textit{split}.
\par
A basis for the real octonion $\mathbb{O}$ will contain eight elements including the identity
\begin{equation}
	1, e_A, \; A=1, \cdots , 7, \; \text{ where } e^2_A=-1.
\label{1-21}
\end{equation}
The elements $e_A$ satisfy the following multiplication table: 
\begin{equation}
	e_A e_B = a_{ABC} e_C - \delta_{AB}
\label{1-22}
\end{equation}
where $a_{ABC}$ is totally antisymmetric and
\begin{equation}
	a_{ABC} = +1 \text{ for } ABC = 123, 516, 624, 435, 471, 673, 572.
\label{1-23}
\end{equation}
For the split octonion algebra we choose the following basis: 
\begin{equation}
\begin{split}
	u_i = &\frac{1}{2} \left( e_i + \imath e_{i+3} \right),	\quad
	u_i^* = \frac{1}{2} \left( e_i - \imath e_{i+3} \right)	, \quad
	i=1,2,3 ;
\\
	u_0 = &\frac{1}{2} \left( 1 + \imath e_7 \right),	\quad
	u_0^* = \frac{1}{2} \left( 1 - \imath e_7 \right)	.
\label{1-30}
\end{split}
\end{equation}
These basis elements satisfy the multiplication table
\begin{align}
	u_i u_j 	&= \epsilon_{ijk} u^*_k,& u^*_i u^*_j &= \epsilon_{ijk} u_k, &i,j,k 		 &= 1,2,3 	&&
\label{1-40} \\
	u_i u^*_j &= - \delta_{ij} u_0, 	&u^*_i u_j 		&= - \delta_{ij} u^*_0,&					 &					&&
\label{1-50}\\
	u_i u_0 	&= 0, 									&u_i u^*_0 		&= u_i	,							 &u^*_i u_0 &= u_i^*	,		&u_i^* u_0^* &=0,
\label{1-60}\\
	u_0 u_i 	&= u_i, 								&u^*_0 u_i 		&= 0	,	  						 &u_0 u^*_i &= 0	,				&u_0^* u_i^* &=u_i^*,
\label{1-70}\\
	u_0^2 	&= u_0,   								&{u^*_0}^2 		&= u^*_0	,	  						 &u_0 u^*_0 &= u_0^* u_0 = 0	.				&&
\label{1-80}
\end{align}
The split octonion algebra contains divisors of zero and hence is not a division algebra.
\par
The algebra of real octonions $\mathbb{O}$ can be represented as
\begin{equation}	
	\mathbb{O} = \mathbb{H} + e_7 \mathbb{H}
\label{1-90}
\end{equation}
where $\mathbb{H}$ denotes the quaternions, spanned by $\left\{ 1, e_1, e_2, e_3 \right\}$;  then basis \eqref{1-30} can be represented as
\begin{equation}
\begin{split}
	u_i = &\frac{e_i}{2} \left( 1 - \imath e_7 \right),	\quad
	u_i^* = \frac{e_i}{2} \left( 1 + \imath e_7 \right)	, \quad
	i=1,2,3 ;
\\
	u_0 = &\frac{1}{2} \left( 1 + \imath e_7 \right),	\quad
	u_0^* = \frac{1}{2} \left( 1 - \imath e_7 \right)	.
\label{1-85}
\end{split}
\end{equation}
A realization of the split octonion algebra is via the Zorn vector matrices
\begin{equation}	
	\left(
	\begin{array}{cc}
		a				&	\vec x	\\
		\vec y	&	b
	\end{array}
	\right)
\label{1-95}
\end{equation}
where $a,b$ are real numbers and $\vec x, \vec y$ are 3-vectors, with the product defined as
\begin{equation}	
	\left(
	\begin{array}{cc}
		a				&	\vec x	\\
		\vec y	&	b
	\end{array}
	\right)
	\left(
	\begin{array}{cc}
		c				&	\vec u	\\
		\vec v	&	d
	\end{array}
	\right) =
	\left(
	\begin{array}{rl}
		ac + \vec x \cdot \vec v										&	\quad a \vec u + d \vec x -
		\vec y \times \vec v 																														 \\
		c \vec y + b \vec v + \vec x \times \vec u	&	\quad bd + \vec y \cdot \vec u
	\end{array}
	\right)
\label{1-100}
\end{equation}
here $( \cdot )$ and $[ \times ]$ denote the usual scalar and vector products.
\par
If the basis vectors of the 3D Euclidean space are $\vec e_i, i=1,2,3$ with
$\vec e_i \times \vec e_j = \epsilon_{ijk} \vec e_k$ and $\vec e_i \cdot \vec e_j = \delta_{ij}$, then we can rewrite the split octonions as matrices
\begin{eqnarray}	
	u_0^* & = & \left(
	\begin{array}{ll}
		1				&	\vec 0	\\
		\vec 0	&	0
	\end{array}
	\right), \quad
	u_i^* = \left(
	\begin{array}{cc}
		0				&	-\vec e_i	\\
		\vec 0	&	0
	\end{array}
	\right), \quad
\label{1-110}\\
	u_0 & = & \left(
	\begin{array}{ll}
		0				&	\vec 0	\\
		\vec 0	&	1
	\end{array}
	\right), \quad
	u_i = \left(
	\begin{array}{cc}
		0				&	\vec 0	\\
		\vec e_i	&	0
	\end{array}
	\right).
\label{1-120}
\end{eqnarray}
The split (and real) octonions are alternative algebras, i.e. for any octonions $a,b$
\begin{equation}	
	\left( a a \right) b = a \left( ab \right), \;
	a \left( bb \right) = \left( ab \right) b, \;
	\left( ab \right) a = a \left( ba \right).
\label{1-130}
\end{equation}

\section{Supersymmetric quantum mechanics}
\label{supersymmetry}

In this section we follow Ref.~\cite{Haymaker:1985us}. A one-dimensional quantum mechanical Hamiltonian
\begin{equation}	
	\hat H = \left(
	\begin{array}{cc}
		\hat H_-	&	0	\\
		0					&	\hat H_+
	\end{array}
	\right)
\label{2-10}
\end{equation}
is said to be supersymmetric \cite{Witten:1981nf} \cite{Cooper:1982dm} if the corresponding potentials $V_\pm (x)$ are related according to 
\begin{equation}	
	V_{\pm} = \frac{{U'}^2}{8} \mp \frac{U''}{4}.
\label{2-20}
\end{equation}
The demonstration that $\hat H$ is supersymmetric hinges on the existence of the generators of supersymmetry $Q, \bar Q$ which together with $\hat H$ satisfy the commutation and anticommutation relations
\begin{eqnarray}	
	\left[ Q,\hat H \right] &=& \left[ \bar Q,\hat H \right] = 0 ,
\label{2-30}\\
	\left\{ \bar Q, \bar Q \right\} &=& \left\{ Q, Q \right\} = 0 ,
\label{2-40}\\
	\left\{ \bar Q, Q \right\} &=& 2 \hat H
\label{2-50}
\end{eqnarray}
here
\begin{eqnarray}	
	Q &=& \left( \hat p - \imath \frac{U'}{2} \right) \sigma^+ ,
\label{2-60}\\
	\bar Q &=& \left( \hat p + \imath \frac{U'}{2} \right) \sigma^- ,
\label{2-70}
\end{eqnarray}
where $\imath^2 = -1$, $\hat p = -\imath \frac{\partial}{\partial x}$ and $\sigma^\pm$ are the 2x2 matrices
\begin{eqnarray}	
	\sigma^- &=& \left(
	\begin{array}{cc}
		0	&	0	\\
		1	&	0
	\end{array}
	\right) ,
\label{2-80}\\
	\sigma^+ &=& \left(
	\begin{array}{cc}
		0	&	1	\\
		0	&	0
	\end{array}
	\right).
\label{2-90}
\end{eqnarray}
Because of the relations $\left\{ \sigma^-, \sigma^+ \right\} = 1$ and
$\left[ \sigma^+, \sigma^- \right] = \sigma_z$, it is easily verified that Eqs.~\eqref{2-30}-\eqref{2-50} are satisfied and that
\begin{equation}	
	\hat H = \frac{1}{2} \left(
		Q \bar Q + \bar Q Q
	\right) =
	\frac{1}{2} \left(
		{\hat p}^2 + \frac{{U'}^2}{4}
	\right) \mathbb I +
	\frac{U''}{4} \sigma_z
\label{2-100}
\end{equation}
where $\mathbb I$ is the identity matrix and $\sigma_z$ is the Pauli matrix.

\section{The nonassociative generalization of supersymmetric quantum mechanics}
\label{extension}

In this section we would like to show that the supersymmetric quantum mechanics has an octonionic generalization.
\par
It is easy to show that $\sigma^-$, $\sigma^+$ and $\sigma_z$ can be identified with the split octonions
\begin{equation}	
	\sigma^- \rightarrow u_1, \;
	\sigma^+ \rightarrow -u_1^*, \;
	\sigma_z \rightarrow u_0^* - u_0.
\label{3-10}
\end{equation}
It suggests to consider the following generalizations of supersymmetric operators $Q, \bar Q$ \eqref{2-60}~\eqref{2-70}
\begin{eqnarray}	
	Q &=& \sum \limits_{i=1}^3 \left( - \hat p_i + \imath V_{,i} \right) u^*_i =
	\sum \limits_{i=1}^3 \mathcal D_i u^*_i ,
\label{3-20}\\
	\bar Q &=& \sum \limits_{i=1}^3 \left( \hat p_i + \imath V_{,i} \right) u_i =
	\sum \limits_{i=1}^3 \bar{\mathcal D}_i u_i.
\label{3-30}
\end{eqnarray}
For constructing the Hamiltonian we need to introduce quadratic operators in $Q$
\begin{eqnarray}	
	Q \bar Q &=& \left[ {\hat p}^2 + \sum \limits_{j=1}^3 \left( V_{,j} \right)^2 +
	\sum \limits_{j=1}^3 V_{,jj} \right] u_0^*,
\label{3-33}\\
	\bar Q Q &=& \left[ {\hat p}^2 + \sum \limits_{j=1}^3 \left( V_{,j} \right)^2 -
	\sum \limits_{j=1}^3 V_{,jj} \right] u_0
\label{3-36}
\end{eqnarray}
here $\hat p_i = - \imath \frac{\partial}{\partial x_i}$,
$V_{,i} = \frac{\partial V}{\partial x_i}$ and
$\sum \limits_{j=1}^3 V_{,jj} = \sum \limits_{j=1}^3 \frac{\partial^2 V}{\partial x_j^2} = \Delta V$. The Hamiltonian is
\begin{equation}	
	\hat H = \frac{1}{2} \left\{ \bar Q, Q \right\} =
	\left(
	\begin{array}{ll}
		\hat H_+	& \vec 0 \\
		\vec 0		& \hat H_-
	\end{array}
	\right) =
	\frac{1}{2} \left[
		{\hat p}^2 + \sum \limits_{j=1}^3 \left( V_{,j} \right)^2
	\right] + \frac{1}{2} \left( - \imath e_7 \right) \sum \limits_{j=1}^3 V_{,jj}
\label{3-50}
\end{equation}
where $- \imath e_7 = u_0^* - u_0$,
$\hat H_+ = \frac{1}{2} \left[ {\hat p}^2 + \sum \limits_{j=1}^3 \left( V_{,j} \right)^2
\right] + \frac{1}{2} \sum \limits_{j=1}^3 V_{,jj}$ and
$\hat H_- = \frac{1}{2} \left[ {\hat p}^2 + \sum \limits_{j=1}^3 \left( V_{,j} \right)^2
\right] - \frac{1}{2} \sum \limits_{j=1}^3 V_{,jj}$. Instead of commutation/anticommutation  relationships \eqref{2-30}-\eqref{2-50} we have
\begin{eqnarray}	
	\left\{ \bar Q, \bar Q \right\} &=& \left\{ Q, Q \right\} = 0 \text{ because of }
	Q^2 = {\bar Q}^2 = 0,
\label{3-60}\\
	\left[ Q, \hat H \right] &=& Q \left( \bar Q Q \right) - \left( Q \bar Q \right) Q =
	\sum \limits_{i,j=1}^3 V_{,ij} \mathcal D_j u^*_i -  Q \sum \limits_{j=1}^3 V_{,jj} ,
\label{3-70}\\
	\left[ \bar Q, \hat H \right] &=& \bar Q \left( Q \bar Q \right) -
	\left( \bar Q Q \right) \bar Q =
	\sum \limits_{i,j=1}^3 V_{,ij} \bar{\mathcal D}_j u_i -
	\bar Q \sum \limits_{j=1}^3 V_{,jj},
\label{3-80}\\
	\left[ Q \bar Q, \hat H \right] &=& \left[ \bar Q Q, \hat H \right] = 0.
\label{3-90}
\end{eqnarray}
Commutations \eqref{3-70} \eqref{3-80} show that the operators $Q, \bar Q, \hat H$ do not constitute a closed algebra because of the appearance of the 
$\sum \limits_{i,j=1}^3 V_{,ij} \bar{\mathcal D}_j u_i$ and
$\sum \limits_{i,j=1}^3 V_{,ij} \bar{\mathcal D}_j u^*_i$ terms. 
\par
Now we can write the Schr\"odinger equation with Hamiltonian \eqref{3-50} which is similar to \eqref{2-100}
\begin{equation}	
	\hat H \Psi =
	\left(
	\begin{array}{ll}
		\hat H_+	& \vec 0 \\
		\vec 0		& \hat H_-
	\end{array}
	\right)
	\left(
	\begin{array}{l}
		\psi_1 \\
		\psi_2
	\end{array}
	\right) =
	E \left(
	\begin{array}{l}
		\psi_1 \\
		\psi_2
	\end{array}
	\right)
\label{3-130}
\end{equation}
here the multiplication rule on the LHS is defined via definition \eqref{3-110} and the Zorn multiplication rule \eqref{1-100}.
\par
The next question is the interpretation of Eq.~\eqref{3-130}. The problem here is that Hamiltonian \eqref{3-50} has the nonassociative number $e_7$, and consequently, such eigenmatrix equations do not always have solutions. It is not clear that the Zorn multiplication rule \eqref{1-100} may give us correct formulation of the eigenvalue problem for any spit octonions. In our case we can avoid these problems in the following way. The operator $- \imath e_7$ has the following Zorn matrix representation: 
\begin{equation}	
	- \imath e_7 = \left(
	\begin{array}{rr}
		1				&	\vec 0	\\
		\vec 0	&	-1
	\end{array}
	\right)
\label{3-100}
\end{equation}
which is similar to Pauli matrix $\sigma_z$. Let us introduce the following notation for a wave function $\Psi$: 
\begin{equation}	
	\Psi = \left(
	\begin{array}{r}
		\psi_1			\\
		\vec \psi_2	
	\end{array}
	\right) = 
	\left(
	\begin{array}{rr}
		\psi_1			&	\vec 0	\\
		\vec \psi_2	&	0
	\end{array}
	\right).
\label{3-110}
\end{equation}
In the same way, we introduce the notation for a Hermitian-conjugated wave function $\Psi^\dag$: 
\begin{equation}	
	\Psi^\dag = \left( 
		\begin{array}{rr}
			\psi_1^*		&	\vec \psi_2^*	
		\end{array}
	\right) = 
	\left(
	\begin{array}{rr}
		\psi_1^*		&	\vec \psi_2^*	\\
		\vec 0			&	0
	\end{array}
	\right) .
\label{3-115}
\end{equation}
The scalar product of the wave functions $\Xi$ and $\Psi$ can be defined as
\begin{equation}	
	\Xi^\dag \Psi =
	\left(
		\begin{array}{rr}
		\xi_1^*	&	\vec \xi_2^* 
		\end{array}
	\right) \left(
	\begin{array}{r}
		\psi_1			\\
		\vec \psi_2	
	\end{array}
	\right) = \left(
	\begin{array}{rr}
		\xi_1^*			&	\vec \xi_2^*	\\
		\vec 0		&	0
	\end{array}
	\right) \left(
	\begin{array}{rr}
		\psi_1			&	\vec 0	\\
		\vec \psi_2	&	0
	\end{array}
	\right) = \left(
	\begin{array}{rr}
		\xi_1^* \psi_1 + \vec \xi_2^* \cdot \vec \psi_2	&	\vec 0	\\
		\vec 0																			 &	0
	\end{array}
	\right) = \xi_1^* \psi_1 + \vec \xi_2^* \cdot \vec \psi_2
\label{3-150}
\end{equation}

One can see that
\begin{equation}	
	- \imath e_7 \left(
	\begin{array}{r}
		\psi_1			\\
		\vec \psi_2	
	\end{array}
	\right) = \left(
	\begin{array}{rr}
		1				&	\vec 0	\\
		\vec 0	&	-1
	\end{array}
	\right) \left(
	\begin{array}{r}
		\psi_1			\\
		\vec \psi_2	
	\end{array}
	\right) = \left(
	\begin{array}{r}
		\psi_1			\\
		- \vec \psi_2	
	\end{array}
	\right)
\label{3-120}
\end{equation}
\par
Now we can discuss the problem of the observability of operators $Q, \bar Q$ and so on. Let us remind that a physical quantity $P$ is observable if the eigenvalue problem for the corresponding operator $\hat P$ has the sense
\begin{equation}	
	\hat P \Psi = P \Psi.
\label{3-160}
\end{equation}
However, relation \eqref{3-120} is not always satisfied for other split octonions, for example, 
\begin{equation}	
	u_i \left(
	\begin{array}{r}
		\psi_1			\\
		\vec \psi_2	
	\end{array}
	\right) = \left(
	\begin{array}{rl}
		0				&	\vec 0	\\
		\vec e_i	&	0
	\end{array}
	\right) \left(
	\begin{array}{r}
		\psi_1			\\
		\vec \psi_2	
	\end{array}
	\right) = \left(
	\begin{array}{rl}
		0									&	- \left[ \vec e_i \times \vec \psi_2 \right]	\\
		\psi_1 \vec e_i 	&	0
	\end{array}
	\right)
\label{3-140}
\end{equation}
and the last matrix cannot be presented as
$\left(
	\begin{array}{r}
		a			\\
		\vec b	
	\end{array}
 \right)
  = \left(
	\begin{array}{rr}
		a				&	0	\\
		\vec b 	&	0
	\end{array}
 \right)
$.
\par
The same is true for
\begin{equation}	
	\left( 
		\begin{array}{rr}
		\xi_1	&	\vec \xi_2 
		\end{array}
	\right) u_i^* =
	\left(
	\begin{array}{rr}
		\xi_1	&	\vec \xi_2	\\
		0			&	0
	\end{array}
	\right)
	\left(
	\begin{array}{rr}
		0				&	- \vec e_i	\\
		\vec 0	&	0
	\end{array}
	\right) =
		\left(
	\begin{array}{rl}
		0																	 				 &	- \xi_1 \vec e_i	\\
		- \left[ \vec \xi_2 \times \vec e_i \right]&	0
	\end{array}
	\right)
\label{3-145}
\end{equation}
and last matrix can not be presented as
$\left(
	a , \vec b
 \right)
  = \left(
	\begin{array}{rr}
		a		&	\vec b	\\
		0 	&	0
	\end{array}
 \right)
$. 
\par
One can say that Zorn matrices \eqref{3-110} and \eqref{3-115} do not form a subalgebra of octonions. It means that for operator $u_i$ the eigenvalue problem does not have any sense. Consequently, the operators $Q, \bar Q$ are \textcolor{red}{\emph{unobservables}}. The operators $Q \bar Q , \bar Q Q, \hat H $ are \textcolor{red}{\emph{observables}}. It confirms the idea presented in  Ref.~\cite{Dzhunushaliev:2007cx} that in general the nonassociative operators (numbers) do not allow us to present these operators as observables of some physical quantities.
\par
Equations \eqref{3-70} and \eqref{3-80} directly show that the operators $Q, \bar Q$ are nonassociative and nonalternative because the associators
\begin{eqnarray}	
	\left( Q \bar Q \right) Q - Q \left( \bar Q Q \right) & \neq & 0,
\label{3-170}\\
	\left( \bar Q Q \right) \bar Q - \bar Q \left( Q \bar Q \right) & \neq & 0
\label{3-180}
\end{eqnarray}
are nonzero.

\section{Unobservables and hidden variables}
\label{hidden}

Commutators \eqref{3-70} and \eqref{3-80} tell us that operators $u_i$ and $u_i^*$ have to be compared to hidden variables of the hidden variables theory. According to \eqref{3-140} we cannot interpret these operators as quantum operators because we cannot determine the action of the operators on wave function \eqref{3-110}. The reason for this is the non-associativity of operators $u_i$ and $u_i^*$: they are split octonions. Let us note that the supersymmetric quantum mechanics presented in Section~\ref{supersymmetry} is usual quantum mechanics becuase  $\sigma^+ = u_1, \sigma^- = u_1^*, \sigma_z = -\imath e_7$ are quaternions that are the associative subalgebra of octonions.
\par
Thus the octonionic generalization of supersymmetric quantum mechanics presented in Section~\ref{extension} is formed with operators $Q$ and $\bar Q$. These operators are built  using split octonions $u_i$ and $u_i^*$. Hamiltonian \eqref{3-50} is constructed in such a way that it has the physical application as the Schr\"odinger equation
\begin{equation}	
	\hat H \Psi = E \Psi .
\label{4-10}
\end{equation}
The special features of this quantum mechanics are as follows:
\begin{enumerate}
	\item the quantities $Q$ and $\bar Q$ are unobservables and 
	\item the quantities $Q \bar Q$ and $\bar Q Q$ are physical observables.
\end{enumerate}
Item (1) means that the Hamiltons equations of motion for the operators $Q$ and $\bar Q$
\begin{eqnarray}
	\frac{d Q}{d t} &=& \imath \left[ \hat H, Q \right] =
	\imath \left( \sum \limits_{i,j=1}^3 V_{,ij} \mathcal D_j u^*_i -
	Q \sum \limits_{j=1}^3 V_{,jj} \right),
\label{4-20}\\
	\frac{d \bar Q}{d t} &=& \imath \left[ \hat H, Q \right] =
	\imath \left( \sum \limits_{i,j=1}^3 V_{,ij} \bar{\mathcal D}_j u_i -
	\bar Q \sum \limits_{j=1}^3 V_{,jj} \right)
\label{4-30}
\end{eqnarray}
have the nonassociative $\sum \limits_{i,j=1}^3 V_{,ij} \mathcal D_j u^*_i$ and
$\sum \limits_{i,j=1}^3 V_{,ij} \bar{\mathcal D}_j u_i$ terms. One can say that quantities $u_i$ and $u_i^*$ are similar to the hidden variables (in the hidden variables theory) in the sense that the Hamilton equations \eqref{4-20} and \eqref{4-30} have  not only terms $\sum \limits_{j=1}^3 V_{,jj} Q$ or
$\sum \limits_{j=1}^3 V_{,jj} \bar Q$ but $\sum \limits_{i,j=1}^3 V_{,ij} \mathcal D_j u^*_i$ and $\sum \limits_{i,j=1}^3 V_{,ij} \bar{\mathcal D}_j u_i$ as well.
\par
This situation should be compared to Bell's theorem. If quantum mechanics has hidden variables
$\vec \lambda \in \Omega \subset \mathbb R^n$ then the probability distribution of the hidden variables in state $\psi$ is $\rho_\psi(\vec \lambda)$. As a probability distribution, $\rho_\psi(\vec \lambda)$ must have the properties
\begin{eqnarray}
	\rho_\psi(\vec \lambda ) & \geq & 0 ,
\label{4-40}\\
	\int \limits_{\mathbb R^n} \rho_\psi(\vec \lambda ) d^n \vec \lambda &=& 1 .
\label{4-50}
\end{eqnarray}
It is important for us that at the proof of Bell's inequalities we connive that hidden variables have a probability distribution. However, the situation for the octonionic generalization of supersymmetric quantum mechanics presented in Section~\ref{extension} is radically different: operators $u_i$ and $u_i^*$ are unobservables. The difference between the unobservables in our case and the hidden variables is that the unobservables are neither classical (because they are nonassociative numbers --- split octonions) nor quantum (because they are unobservables) variables. The special peculiarity of unobservables is that they are nonassociative quantities.
\par 
Thus the nonassociative unobservables \textcolor{red}{\emph{do not provide a way to violate Bell's inequalities.}} This takes place because the hidden variables in the theory of hidden variables \textcolor{red}{\emph{can be measured in principle.}} However, the unobservables  presented here \textcolor{red}{\emph{cannot be measured in principle.}}

\section{Physical applications}
\label{application}

The above mentioned consideration shows that the octonionic generalization of supersymmetric quantum mechanics describes an observable particle formed from unobservable ``particles''.
\par
Let us note that a similar idea about unobservable variables existing in the $t$-$J$ model with High-T$_c$ superconductivity. It is a widely spread opinion that the low energy physics of High-T$_c$ cuprates is described in terms of $t$-$J$ type model, which is given by \cite{LN9221}
\begin{equation}
	H = \sum \limits_{i,j} J\left(
	{{S}}_{i}\cdot {{S}}_{j}-\frac{1}{4} n_{i} n_{j} \right)
	-\sum_{i,j} t_{ij}
	\left(c_{i\sigma}^\dagger
	c_{j\sigma}+{\rm H.c.}\right)
\label{4c-30}
\end{equation}
where $t_{ij}=t$, $t'$, $t''$ for the nearest, second nearest and 3rd nearest neighbor pairs, respectively. In this model the electron operator is presented as
\begin{equation}
	c^\dagger_{i\sigma} = f_{i\sigma}^\dagger b_{i}
\label{4c-50}
\end{equation}
where $f_{i\sigma}^\dagger$, $f_{i \sigma}$ are the fermion operators, while $b_{i}$ is the
slave-boson operator. This representation together with the constraint
\begin{equation}
	f_{i\uparrow}^\dagger f_{i\uparrow} + f_{i\downarrow}^\dagger f_{i\downarrow} +
	b^\dagger_{i} b_{i} = 1
\label{4c-60}
\end{equation}
reproduces all the algebra of the electron operators. The physical meaning of the operators $f$ and $b$ is unclear: do  these fields exist or not ?
\par
If we compare factorization \eqref{4c-50} and operators $Q, \bar Q$ and $Q \bar Q, \bar Q Q$ we can presuppose that the operators $f_{i\sigma}^\dagger, b_{i}$ are elements of an infinite dimensional nonassociative algebra $\mathfrak Q$. This algebra has an associative subalgebra
$\mathfrak A \subset \mathfrak Q$ and the operator $c^\dagger_{i\sigma} \in \mathfrak A$ is observable but the operators $f_{i\sigma}^\dagger, b_{i} \in \mathbb Q \backslash \mathfrak A$ are unobservables. It could mean that the High-T$_c$ superconductivity (similar to quantum chromodynamics) can be understood on the basis of a \emph{nonperturbative} quantum theory and one can assume that the  non-perturbative quantum theory (on the operator language) could be realized as a nonassociative quantum theory (realized as a nonassociative algebra
$\mathfrak Q$) with observables belonging to an associative subalgebra $\mathfrak A$ and unobservables belonging to $\mathfrak Q \backslash \mathfrak A$.
\par
It is necessary to note here that in Ref's~\cite{Niemi:2005qs}-\cite{oma1} there is a classical generalization of slave-boson decomposition on gauge theories, which is so called ``spin-charge separation''.
\par
The next question naturally appearing here is as follows: is it possible to apply the idea presented here to the description of the unobservability of quarks in quantum chromodynamics ?

\section{Discussion and conclusions}
\label{conclusion}

Hidden variable theories were espoused by a minority of physicists who argued that the statistical nature of quantum mechanics indicated that quantum mechanics is "incomplete". In quantum mechanics, the question arises whether there might be some deeper reality hidden beneath quantum mechanics, which is to be described by a more fundamental theory that can always predict the outcome of each measurement with certainty. A minority of physicists maintain that underlying the probabilistic nature of the universe is an objective foundation/property -- the hidden variable. The main point of the hidden variables in quantum mechanics is that they can describe the movement of a quantum particle in a deterministic manner. For us such description seems a nondeterministic one because an underlying hidden variables theory is too complicated. It is very important that for the hidden variables, we can introduce a probability distribution $\rho_\psi(\vec \lambda)$ describing the hidden variables in a state $\psi$. The situation for the unobservables in a nonassociative quantum theory is radically different: the unobservables cannot have any probability distribution in principle. 
\par 
For the octonionic generalization of the supersymmetric quantum mechanics presented here, the situation is similar: the observables are the quantities which are presented as a multilinear combination of more elementary quantities -- unobservables. For instance the Hamiltonian 
$\hat H$ \eqref{3-50} is the bilinear combination of unobservables $Q$ and $\bar Q$. In contrast to the hidden variables theory, the unobservables in the octonionic generalization of the supersymmetric quantum mechanics are unobservables in principle, i.e. for these unobservables we cannot assign any probability distribution \eqref{4-40}. According \cite{jordan} and \cite{emch} octonion valued observables become admissible only in the case of three degrees of freedom. Octonion valued fields with an infinite number of degrees of freedom can only operate in a nonobservable Hilbert space. 
\par 
Let us note that the unobservability idea was applied in Ref.~\cite{Gunaydin:1974} for the description of the unobservability of quark states and observability of mesons and nucleons. The idea presented there is to describe quarks (and their associated color gauge bosons) in an octonionic Hilbert space. States in such a space will not all be observable because the propositional calculus of observable states as developed by Birkhoff and von Neumann \cite{birkhoff} can only have realizations as projective geometries corresponding to Hilbert spaces over associative composition algebras, while octonions are nonassociative. In Ref.~\cite{Gunaydin:1974} an observable subspace arises in the following way: within Fock space there will be states which are observable (longitudinal, in the notation of Ref.~\cite{Gunaydin:1974}) which are the linear combinations of $u_0$ and $u^*_0$. The states in transversal direction (spanned by $u_i$ and $u_i^*$) are unobservables. Let us note that the octonionic generalization of the supersymmetric quantum mechanics presented here is based on a  similar consideration for $u_0, u_0^*$ and $u_i, u_i^*$. 
\par
We have shown that supersymmetric quantum mechanics has a split octonionic generalization. The algebra of variables in the octonionic generalization of supersymmetric quantum mechanics is split into observables and unobservables. The unobservables in some sense are similar to hidden varibles but they cannot be measured in principle.

\end{document}